\documentclass[prd,aps,twocolumn,a4paper,floatfix,amsmath,amssymb,nofootinbib]{revtex4-1}
\usepackage{graphicx,color}
\newcommand{\bp}{{\bar p}}
\newcommand{\bq}{{\bar q}}
\newcommand{\order}{{\mathcal O}}

\begin{document}

\title{Critical gravitational collapse with angular momentum}

\author{Carsten Gundlach}

\affiliation{Mathematical Sciences,
University of Southampton, Southampton SO17 1BJ, United Kingdom}

\author{Thomas W. Baumgarte}

\affiliation{Department of Physics and Astronomy, Bowdoin College,
  Brunswick, ME 04011, USA}

\date{22 September 2016}

%%%%%%%%%%%%%%%%%%%%%%%%%%%%%%%%%%%%%%%%%%%%%%%%%%%%%%%%%%%%%%%%%%%%%%%%

\begin{abstract}

We derive a theoretical model of mass and angular momentum scaling in
type-II critical collapse with rotation. We focus on the case where
the critical solution has precisely one, spherically symmetric,
unstable mode. We demonstrate agreement with numerical results for
critical collapse of a rotating radiation fluid, which falls into this
case.

\end{abstract}

%%%%%%%%%%%%%%%%%%%%%%%%%%%%%%%%%%%%%%%%%%%%%%%%%%%%%%%%%%%%%%%%%%%%%%%%

\maketitle

%%%%%%%%%%%%%%%%%%%%%%%%%%%%%%%%%%%%%%%%%%%%%%%%%%%%%%%%%%%%%%%%%%%%%%%%

\section{Introduction}
\label{section:introduction}

%%%%%%%%%%%%%%%%%%%%%%%%%%%%%%%%%%%%%%%%%%%%%%%%%%%%%%%%%%%%%%%%%%%%%%%%

Critical collapse in general relativity refers to phenomena that occur
at the threshold, in the space of initial data, between data that lead
to black hole formation (collapse) and those that do not.  Regular
initial data can be classified as supercritical or subcritical
according to whether or not they form a black hole.  We refer to the
boundary between supercritical and subcritical data as the black-hole
threshold, or the critical surface. In type-II critical collapse, the
black-hole mass formed by supercritical data becomes arbitrarily small
as the threshold is approached, and scales as a universal power of
distance from this threshold.  The exponent in these power laws is
referred to as the critical exponent.  Critical collapse was first
reported in the seminal work of Choptuik \cite{Choptuik1993}, who
performed numerical time evolutions of a massless scalar field in
spherical symmetry. Soon afterwards, similar results were reported for
a radiation fluid, i.e.~a perfect fluid with the ultra-relativistic
equation of state $P=\rho/3$ \cite{EvansColeman} (where $P$ is the
pressure and $\rho$ the total energy density), and for axisymmetric
gravitational waves in vacuum \cite{AbrahamsEvans}.  The literature on
numerous further numerical experiments as well as theoretical
derivations of the scaling laws is reviewed in \cite{GundlachLRR}.

In \cite{critfluidpert}, one of us (CG) showed that the spherically
symmetric, continuously self-similar critical solution for perfect
fluid collapse with the equation of state $P=\kappa\rho$ has only a
single ($l = 0$) unstable mode for the range
$1/9<\kappa\lesssim 0.49$, which includes radiation fluids with
$\kappa = 1/3$.  Based on this, and the more general theory given in
\cite{angmom}, CG predicted power-law scaling for the black-hole mass
and angular momentum for initial data with small deviations from
spherical symmetry, and computed numerical values for the critical
exponents.  

In \cite{BaumgarteMontero}, the other one of us (TWB), together with
Montero, carried out the first critical collapse simulations of a
radiation fluid in the absence of spherical symmetry.  More recently,
we generalized these simulations to study critical collapse with
angular momentum \cite{BaumgarteGundlach}.  Specifically, we
considered a two-parameter family of initial data describing rotating
radiation fluids, with one parameter $\eta$ controlling the strength
of the initial data and a second parameter $\Omega$ their angular
momentum.  These simulations confirmed the critical exponents found in
\cite{angmom} and provided evidence for their universality.

In Sec.~\ref{sec2} we provide a
self-contained derivation of the scaling laws in rotating critical
collapse, and in Sec.~\ref{sec3} we demonstrate agreement
with the numerical results of \cite{BaumgarteGundlach} for radiation
fluids with $\kappa=1/3$.  Sec.~\ref{sec4} contains a summary and
discussion of our results.

%%%%%%%%%%%%%%%%%%%%%%%%%%%%%%%%%%%%%%%%%%%%%%%%%%%%%%%%%%%%%%%%%%%%%%%%

\section{Scaling laws for rotating critical collapse}
\label{sec2}

%%%%%%%%%%%%%%%%%%%%%%%%%%%%%%%%%%%%%%%%%%%%%%%%%%%%%%%%%%%%%%%%%%%%%%%%

Consider an analytic family of regular initial data parameterized by
two parameters $p$ and ${\bf q}$.  We assume that, if these data
evolve to form a black hole, the black-hole mass $M$ and angular
momentum ${\bf J}$ obey the symmetries
\begin{subequations} \label{symm}
\begin{eqnarray}
\label{psymm}
M(p,- {\bf q})&=&M(p,{\bf q}), \\
\label{qsymm}
{\bf J}(p,-{\bf q})&=&-{\bf J}(p,{\bf q}).
\end{eqnarray}
\end{subequations}
A sufficient condition for these two assumptions to hold is that ${\bf
  q} \to - {\bf q}$ corresponds to a spatial reflection of the initial
data.  The assumption (\ref{qsymm}) implies that initial data with
${\bf q}=0$ form a non-spinning black hole, but not that they are
necessarily spherically symmetric. In the following, for simplicity of
notation, we restrict to axisymmetry, so that $q$ and $J$ become
numbers.  

The black hole threshold within such a two-parameter family is a curve
in the $(p,q)$-plane that is symmetric under $q\to-q$. We can
fine-tune the initial data to the black hole threshold, in practice by
bisection along any smooth 1-parameter family of initial data that
crosses it.

The general theory of type-II critical collapse \cite{GundlachLRR} is
based on the assumption that the solution first evolves towards an
intermediate regime, during which it is approximated by a
universal critical solution that contracts in a self-similar fashion.
The critical solution has at least one unstable mode, however, which
ultimately drives the evolution either towards black-hole formation or
dispersal.  In the following we discuss these different stages of
the time evolution separately.

%%%%%%%%%%%%%%%%%%%%%%%%%%%%%%%%%%%%%%%%%%%%%%%%%%%%%%%%%%%%%%%%%%%%%%%%

\subsection{From the initial data to the intermediate self-similar regime}
\label{sec2a}

%%%%%%%%%%%%%%%%%%%%%%%%%%%%%%%%%%%%%%%%%%%%%%%%%%%%%%%%%%%%%%%%%%%%%%%%

According to our assumption, initial data sufficiently close to the
black-hole threshold evolve to an intermediate regime during
which the solution $Z$ is given by a critical solution $Z_*$ plus
linear perturbations. Here $Z$ denotes a set of first-order in time
dynamical variables and $({\bf x},\tau)$ a set of $3+1$
coordinates, such that $Z({\bf x},\tau)=Z({\bf x})$ if and only
if the spacetime is continuously self-similar (homothetic).

Variables and coordinates $Z({\bf x},\tau)$ can be constructed as
follows. If we write the physical spacetime metric $g_{ab}$ in the
adapted coordinates $({\bf x},\tau)$ as
\begin{equation}
g_{\mu\nu}({\bf x},\tau)=e^{-2\tau}\bar g_{\mu\nu}({\bf x},\tau),
\end{equation}
then the spacetime is continuously self-similar with the homothetic
vector $\partial/\partial \tau$ if and only if all components of the
conformal metric $\bar g_{\mu\nu}({\bf x},\tau)$ are
independent of $\tau$.  If we now choose the surfaces of constant
$\tau$ to be spacelike, then $\tau$ is both a time coordinate and the
logarithm of overall spacetime scale. In spherical symmetry or
axisymmetry, a natural choice for the remaining coordinates would be a
rescaled radius $x=e^\tau r$, where $r$ is, for example, an areal
radius, plus two angles. We could then carry out the usual 3+1 split
of $\bar g_{\mu\nu}$. Assuming for simplicity that the lapse and shift
are evolved using first-order in time coordinate conditions, and that
the matter is a perfect fluid with the simple linear
(ultrarelativistic) equation of state $P=\kappa\rho$, (as is the case
for our simulations in \cite{BaumgarteGundlach}), we could choose the
variables $Z$ to be the first-order metric variables $\bar g_{ij}$,
$\bar K_{ij}$, $\bar \alpha$, $\bar\beta^i$, the fluid 3-velocity
$v^i$, and $\bar\rho := e^{2\tau}\rho$. The initial data $Z({\bf
  x},\tau_0)$ for the barred quantities define the solution only up to
an overall scale. This scale is given by $e^{-\tau_0}$. (More
precisely, it is $L e^{-\tau_0}$, where $L$ is an arbitrary constant
of dimension length, in units where $c=G=1$, but for simplicity of
notation we choose units where $L=1$.) Note that we do not need to use
these coordinates in numerical time evolutions.

%% , or even to construct them in post-processing, unless we want to
%% verify self-similarity explicitly.

Using this notation, we can now write the intermediate, approximately
self-similar regime as
\begin{eqnarray}
\label{ansatz}
Z({\bf x},\tau) &\simeq & Z_*({\bf x}) + P(p,q)\, e^{\lambda_0\tau}Z_0({\bf x})
\nonumber \\ &&
+ \, Q(p,q)\,e^{\lambda_1\tau}Z_1({\bf x}) \nonumber \\ &&
+ \, \hbox{decaying perturbations}. 
\end{eqnarray}
Here, $Z_*({\bf x})$ is the critical solution, which, in perfect fluid
critical collapse, is both continuously self-similar and spherically
symmetric.\footnote{For scalar fields, $Z_*$ is discretely
  self-similar and spherically symmetric, and for vacuum gravity it is
  believed to be discretely self-similar and axisymmetric. These
  symmetries are more complicated on a technical level, but the basic
  ideas presented here are unchanged.}  $Z_0$ is the unique growing
spherical mode ($l = 0$), and so $\lambda_0 > 0$.  The amplitude of
this mode, $P(p,q)$, depends on the parameters $p$ and $q$ of the
initial data. $Z_1$ is an $l =1$ axial mode, namely either the unique
growing one (in which case we have $\lambda_1 > 0$), or the least
damped one ($\lambda_1 \leq 0$). Its amplitude $Q(q,p)$ again depends
on the initial data. We normalise $Z_0$ and $Z_1$ later. When there
are two growing modes, we single out both in the analysis because they
dominate the dynamics. When only $Z_0$ is growing, we still need to
keep track of $Z_1$ because it is closely linked to black hole angular
momentum: Kerr is an axial $l =1$ perturbation of Schwarzschild to
linear order in $J/M^2$ \cite{angmom}.

We now define the specific moment of time
\begin{equation} \label{def_tau_star}
\tau_* := -{1\over\lambda_0}\ln|P|,
\end{equation}
which we assume to be in the intermediate self-similar regime.  At
$\tau=\tau_*$, the lengthscale of the solution is then
given by
\begin{equation} \label{scale}
e^{-\tau_*} = |P|^{1/\lambda_0},
\end{equation} 
and, since
$|P|e^{\lambda_0\tau_*} = 1$, 
we have the intermediate Cauchy data
\begin{eqnarray}
\label{data}
Z({\bf x},\tau_*) &\simeq& Z_*({\bf x}) \pm Z_0({\bf x}) + \delta
Z_1({\bf x}) \nonumber \\ && +\hbox{decaying perturbations}.
\end{eqnarray}
Here the sign in front of $Z_0$ is that of $P$; it appears
because of the absolute value taken in the definition
(\ref{def_tau_star}). We have also defined
\begin{equation}
\label{definedelta}
\delta := Q |P|^{-\epsilon},
\end{equation}
with
\begin{equation} \label{defineepsilon}
 \epsilon := {\lambda_1\over\lambda_0}.
\end{equation}

For any 2-parameter family of initial data with parameters $(p,q)$
that obey the symmetries (\ref{symm}) we define the black hole
threshold, i.e.~the critical curve separating supercritical data from
subcritical data, by $(p,q)=(p_*,q_*)=(p_{\rm crit}(q_*),q_*)$. We
also define $p_{*0} := p_{\rm crit}(0)$, {and shall refer to
  $(p,q)=(p_{*0},0)$ as the critical point}. From the symmetry
(\ref{qsymm}), $p_{\rm crit}(q)=p_{\rm crit}(-q)$. To fix the
coordinate freedom on the space of initial data to first order about
the critical point [by linear transformations of $p$ and $q$ that
  respect (\ref{symm})], we define the ``reduced parameters''
\begin{equation}
\label{reduced}
\bp := C_0 (p-p_{*0}),
\qquad
\bq := C_1 q,
\end{equation}
where $C_0$ and $C_1$ are family-dependent constants. They will be
fixed later. 

If $P(p,q)$ and $Q(p,q)$ are analytic (because the initial data are
analytic), we can expand them in powers of $\bp$ and $\bq$. By
definition $P$ vanishes at the critical point $\bp=\bq=0$.  Moreover,
from the symmetry (\ref{psymm}), $P$ must be even in $\bq$. This
suggests that we treat $\bp$ and $\bq^2$ as the same order of
smallness when expanding about the critical point. 
From the symmetry (\ref{qsymm}), $Q$ must be odd
in $\bq$. We may therefore expand
\begin{subequations} \label{PQdef}
\begin{eqnarray}
\label{Pdef}
P&=&\bp-K\bq^2+\order(\bp^2,\bp\bq^2,\bq^4), \\
\label{Qdef}
Q&=&\bq+\order(\bq^3,\bp\bq),
\end{eqnarray}
\end{subequations}
where we have now fixed the family-dependent constants $C_0$ and $C_1$
so that the leading-order terms in these expansions have coefficients
of unity.

The coefficient $K$, and the coefficients of the higher-order terms
that we have not written out here, also depend on the two-parameter
family of initial data because they depend on the nonlinear and
non-universal evolution from generic initial data to the universal
intermediate regime (\ref{ansatz}). We have inserted the minus sign in
(\ref{Pdef}) as we anticipate that $K$ will then be positive: spin
should resist collapse. Since the critical surface corresponds to $P =
0$, the expansion (\ref{Pdef}) also implies that, to leading order,
the critical surface forms a parabola in the $(p,q)$-plane.

%%%%%%%%%%%%%%%%%%%%%%%%%%%%%%%%%%%%%%%%%%%%%%%%%%%%%%%%%%%%%%%%%%%%%%%%

\subsection{From the intermediate self-similar regime to the final
  black hole}

%%%%%%%%%%%%%%%%%%%%%%%%%%%%%%%%%%%%%%%%%%%%%%%%%%%%%%%%%%%%%%%%%%%%%%%%

The key observation for scaling is the following.  As discussed above,
initial data sufficiently close to the black-hole threshold pass
through an intermediate self-similar phase.  During this phase, we can
identify the intermediate Cauchy data (\ref{data}) at $\tau = \tau_*$.
These constitute two universal 1-parameter families of scale-invariant
intermediate initial data, parameterised by the sign $\pm$ of $P$ and
the parameter $\delta$.  The scale-invariant data are completed by the
overall length scale $e^{-\tau_*}$. 

Because the evolution equations are scale-invariant $e^{-\tau_*}$,
translates into an overall length and time scale of the
solution at all subsequent times.  This means that if a
feature of the subsequent spacetime evolution has dimension $L^n$
(where $L$ denotes length, in units where $c=G=1$), then this feature
must be proportional to $e^{-n \tau_*}$.

In particular, $M$ must be proportional to the overall scale
$e^{-\tau_*} = |P|^{1/\lambda_0}$ of the Cauchy data (\ref{data}).
Moreover, the constant of proportionality can depend only on the sign
$\pm$ of $P$ and the dimensionless number $\delta$.  We may therefore
express it in terms of two functions $F^\pm_M(\delta)$. Similarly, $J$
must be proportional to $e^{-2\tau_*} = |P|^{2/\lambda_0}$, and again
the constant of proportionality can be expressed in terms of two
functions of $F^\pm_J(\delta)$.  With
\begin{equation}
\gamma_M := {1\over\lambda_0}
\end{equation}
we therefore have \cite{scalingfunctions}
\begin{subequations} \label{scaling1}
\begin{eqnarray}
\label{Mscaling1}
M(p,q)& \simeq& |P|^{\gamma_M} F_M^\pm(\delta), \\
\label{Jscaling1}
J(p,q)& \simeq& |P|^{2\gamma_M} F_J^\pm(\delta).
\end{eqnarray}
\end{subequations}
We note that the dimensionless quantity $J/M^2$ can only depend on the
sign $\pm$ and $\delta$. Numerical evidence also shows that collapse
actually happens only for $P>0$, and so in the following we ignore the
functions $F^-_{M,J}$, and write $F_{M,J}$ short for $F_{M,J}^+$.

Because of the symmetries (\ref{symm}), $F_M(\delta)$ is even in
$\delta$ and $F_J(\delta)$ is odd. We can now normalize $Z_0$ and
$Z_1$ in such a way that that, to leading order, as $\delta
\rightarrow 0$, we have
\begin{equation}
\label{trivialscalingfunctions}
F_M(\delta)= 1+\order(\delta^2), \qquad F_J(\delta)=\delta+\order(\delta^3).
\end{equation}
In particular, Eqs.~(\ref{scaling1}) then imply that
\begin{equation} \label{JM2}
J/M^2 = \delta+\order(\delta^3).
\end{equation}

Inserting the leading-order expressions
(\ref{trivialscalingfunctions}) together with the definitions (\ref{definedelta}) and (\ref{defineepsilon}) into (\ref{scaling1}) yields
\begin{subequations} \label{scaling2}
\begin{eqnarray}
\label{Mscaling2}
M &\simeq& P^{\gamma_M}, \\
\label{Jscaling2}
J &\simeq& P^{\gamma_J}Q,
\end{eqnarray}
\end{subequations}
where we have also defined
\begin{equation}
\gamma_J := {2-\lambda_1\over\lambda_0}.
\end{equation}
In geometric terms, $P$ and $Q$ are locally smooth scalar functions on
the manifold of smooth initial data, such that $P=0$ gives the
critical surface, and $M$ and $J$ given by (\ref{scaling2}) are
non-smooth at the critical surface precisely because of the
non-integer powers $\gamma_M$ and $\gamma_J$. A related observation is
that the power-law scalings of $M$ and $J$ at the collapse threshold
show the same powers for any 1-parameter family of initial data that
crosses the threshold, independently of the angle at which the
threshold is crossed (compare Table~I in \cite{BaumgarteGundlach}). If
we further use the lowest-order approximations for $P$ and $Q$,
Eqs.~(\ref{PQdef}), we obtain
\begin{subequations} \label{scaling3}
\begin{eqnarray}
\label{Mscaling3}
M &\simeq& (\bp-K\bq^2)^{\gamma_M}, \\
\label{Jscaling3}
J &\simeq& (\bp-K\bq^2)^{\gamma_J}\bq. 
\end{eqnarray}
\end{subequations}

Whether or not the leading-order expressions
(\ref{trivialscalingfunctions}) are an adequate approximation for the
scaling functions $F_{J,M}$ depends on whether the critical solution
has one or two growing modes. In the case of one growing mode,
Eq.~(\ref{defineepsilon}) gives $\epsilon < 0$ and so from 
(\ref{definedelta}) $\delta \rightarrow 0$ as the black-hole
threshold $P = 0$ is approached.  
In the case of two growing modes, however, we have $\epsilon > 0$, so
that we expect large values of $\delta$ to be explored close to the
black-hole threshold.  

%%%%%%%%%%%%%%%%%%%%%%%%%%%%%%%%%%%%%%%%%%%%%%%%%%%%%%%%%%%%%%%%%%%%%%%%

\section{Comparison with numerical experiments}
\label{sec3}

%%%%%%%%%%%%%%%%%%%%%%%%%%%%%%%%%%%%%%%%%%%%%%%%%%%%%%%%%%%%%%%%%%%%%%%%

The expressions (\ref{scaling3}) are the main result of this paper for
the case where the critical solution has only one growing mode. In the
following we compare these predictions with results from numerical
time evolutions for rotating radiation fluids, i.e.~a perfect fluid
with equation of state $P = \kappa \rho$ and $\kappa = 1/3$, for which
there is only a single growing mode.

%%%%%%%%%%%%%%%%%%%%%%%%%%%%%%%%%%%%%%%%%%%%%%%%%%%%%%%%%%%%%%%%%%%%%%%%

\subsection{Numerical setup}

%%%%%%%%%%%%%%%%%%%%%%%%%%%%%%%%%%%%%%%%%%%%%%%%%%%%%%%%%%%%%%%%%%%%%%%%

We consider the 2-parameter family of initial data previously
presented in \cite{BaumgarteGundlach}.  {Specifically, the initial
  density distribution $\rho$ is a Gaussian}.  The overall strength of
this density distribution is parametrized by $\eta$, and its angular
momentum scales with $\Omega$ (see Eqs.~(6) and (7) in
\cite{BaumgarteGundlach}).  For $\Omega = 0$ our initial data reduce
to those considered in \cite{EvansColeman}. The parameters $\eta$ and
$\Omega$ are instances of the generic parameters $p$ and $q$ in
Sect.~\ref{sec2}.  We evolve these data with a code that has been
described in \cite{Baumgarteetal13,Baumgarteetal15,BaumgarteMontero}.
Briefly summarized, we solve the Baumgarte-Shapiro-Shibata-Nakamura
formulation of Einstein's equations
\cite{Nakamura,ShibataNakamura,BaumgarteShapiro} in spherical polar
coordinates, using moving-puncture gauge conditions.  Details of our
numerical specifications can be found in
\cite{BaumgarteMontero,BaumgarteGundlach}.

For initial data that form a black hole we locate a marginally
outermost trapped surface, or apparent horizon, using the technique
described in \cite{ShibataUryu}, and measure its irreducible mass
$M_{\rm irr}$ and angular momentum $J$ as in \cite{Dreyer}.  Both mass
and angular momentum increase a little after an apparent horizon is
first formed, as the black hole accretes some more material from the
surrounding fluid, but they soon settle down to equilibrium values.
Assuming that the new black hole is a Kerr black hole, we then compute
the Kerr mass $M= M_{\rm irr} \left( 1 + (J/M_{\rm
  irr}^2)^2/4\right)^{1/2}$ from the equilibrium values of $M_{\rm
  irr}$ and $J$. Note that our theoretical derivation of the scaling
laws applies equally to $M_{\rm irr}$ and $M$. 

%%%%%%%%%%%%%%%%%%%%%%%%%%%%%%%%%%%%%%%%%%%%%%%%%%%%%%%%%%%%%%%%%%%%%%%%

\subsection{Self-similarity}

%%%%%%%%%%%%%%%%%%%%%%%%%%%%%%%%%%%%%%%%%%%%%%%%%%%%%%%%%%%%%%%%%%%%%%%%

%%%%%%%%%%%%%%%%%%%%%%%%%%%%%%%%%%%%%%%%%%%%%%%%%%%%%%%%%%%%%%%%%
\begin{figure}
% Figure produced with commant self_crit in self_crit.py in /mnt/research/tbaumgar/Work/RotRadFluid/Rot
\includegraphics[width=3in]{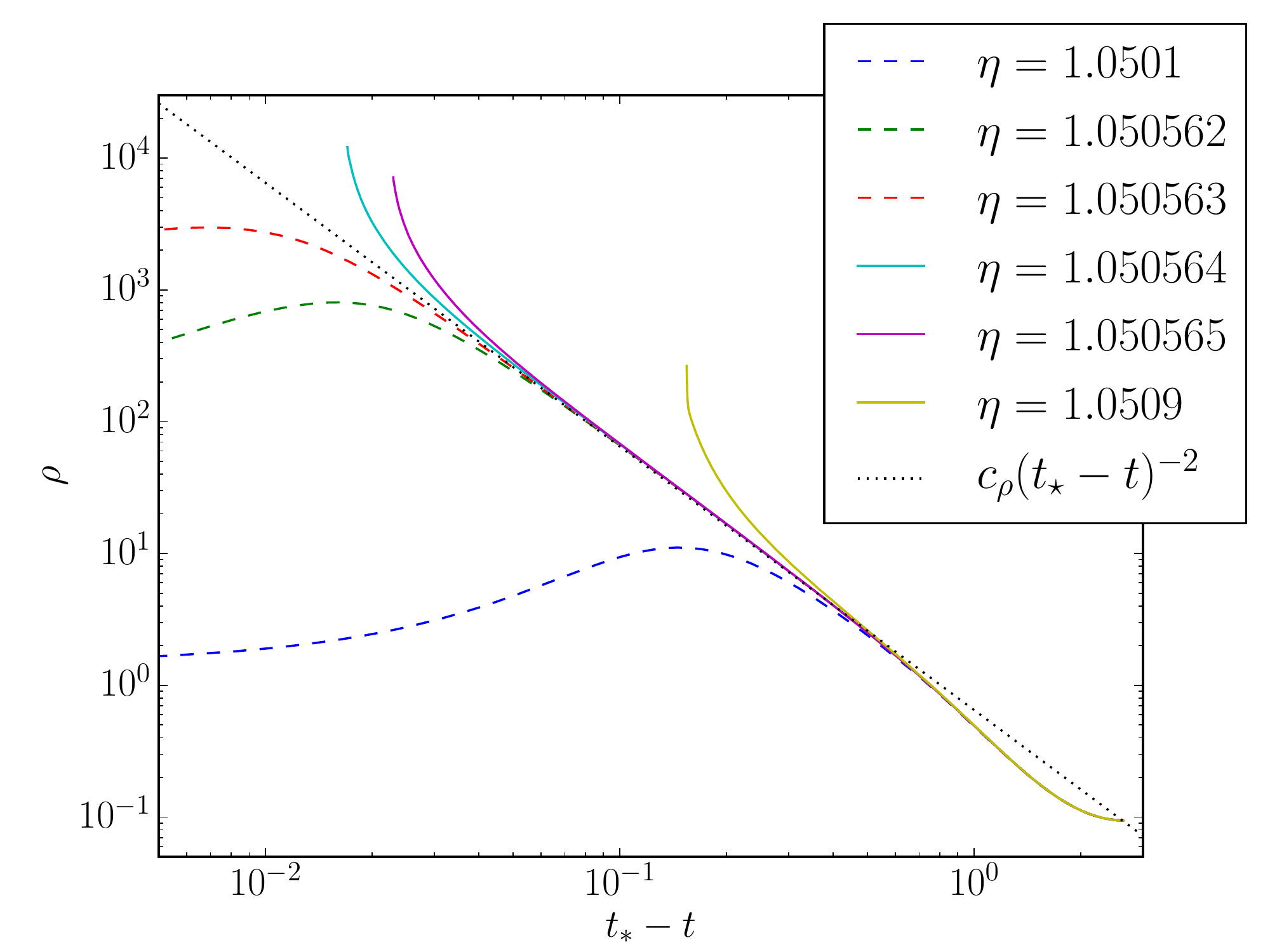} 
\caption{The central density $\rho$ versus $t_* - t$ for several
  evolutions with $\Omega = 0.3$.  Here $t$ is the proper time as
  measured by an observer at the center, and $t_*$ is the accumulation
  time.  Supercritical evolutions are marked
  by sold lines, and subcritical evolutions by dashed lines.  The
  dotted line marks the expression (\ref{rho}) with $c_\rho = 0.65$
  and $t_* = 2.6395$ for the critical solution.}
\label{Fig:self_sim}
\end{figure}
%%%%%%%%%%%%%%%%%%%%%%%%%%%%%%%%%%%%%%%%%%%%%%%%%%%%%%%%%%%%%%%%%

We start by presenting evidence of the underlying assumption of
Sect.~\ref{sec2}, namely that sufficiently close to the black-hole
threshold the evolution passes through an intermediate self-similar
phase.  The evolution during the self-similar phase is governed by
only one length-scale.  On dimensional grounds the density $\rho$ at
the center of symmetry must therefore scale with
\begin{equation} \label{rho}
\rho \simeq c_\rho (t_* - t)^{-2},
\end{equation}
where $t$ is the proper time measured by an observer at the center and
$t_*$ is the accumulation time of the self-similar solution.

In Fig.~\ref{Fig:self_sim} we plot $\rho$ versus $t_* - t$ for several
members of our family initial data with $\Omega = 0.3$, for which the
critical value of $\eta$ is approximately $\eta_* \simeq 1.0505635$.
We include (\ref{rho}) with $c_\rho = 0.65$ and $t_* = 2.6395$ as a
fit for the critical solution.  All evolutions start in the lower
right, and evolve towards the top left.  By coincidence, the
evolutions start from a point that is close to the dotted line marking
the critical solution, but then move away before joining it for real
at $t_* - t \simeq 0.5$. After that, evolutions with initial data closer to
the threshold remain close to the critical solution for a longer time.
In Fig.~\ref{Fig:self_sim} solid lines mark supercritical evolutions,
for which the density ultimately diverges as a black hole is formed,
while dashed lines mark subcritical evolutions, for which the density
ultimately drops to zero as the fluid disperses to infinity.  The
evolutions for $\eta = 1.050563$ and 1.050564 bracket the critical
solution, and follow the central density as given by (\ref{rho}) over
more than two orders of magnitude.  However, even for relatively small
deviations of $\eta$ from the critical value, e.g.~$\eta = 1.0501$ or
1.0509 in Fig.~\ref{Fig:self_sim}, the solution does not appear to go
through a phase of self-similar contraction at all.

%%%%%%%%%%%%%%%%%%%%%%%%%%%%%%%%%%%%%%%%%%%%%%%%%%%%%%%%%%%%%%%%%%%%%%%%

\subsection{Power-law scalings}

%%%%%%%%%%%%%%%%%%%%%%%%%%%%%%%%%%%%%%%%%%%%%%%%%%%%%%%%%%%%%%%%%%%%%%%%

We start our analysis by considering the same 1-parameter families of
data that we previously considered in \cite{BaumgarteGundlach}, namely
sequences for $\Omega = 0$, 0.05, 0.1 and 0.3, as well as for $\eta =
1.02$, 1.035 and 1.0505.  All families cross the critical curve;
locating these intersections provides points $(\eta_*,\Omega_*)$ on
the collapse threshold. Our initial observation is that the location
of these six points is well approximated by a parabola, as expected
from (\ref{Pdef}); see Fig.~1 of \cite{BaumgarteGundlach}.

For supercritical data we can plot the logarithm of the black-hole
masses $M$ and angular momenta $J$ versus the logarithm of the
distance from the critical parameter (see Fig.~2 in
\cite{BaumgarteGundlach}).  Measuring the slope of the resulting
straight lines then provides a numerical estimate of the critical
exponents $\gamma_M$ and $\gamma_J$.  Numerical data can be found in
Table I of \cite{BaumgarteGundlach}. For all sequences considered the
numerical values are within a few percent of the analytical values
\begin{equation}
\gamma_M\simeq 0.3558, \qquad \gamma_J={5\over 2}\gamma_M\simeq 0.8895
\end{equation}
for $\kappa = 1/3$, as computed by \cite{Maison} and
\cite{critfluidpert}, respectively.  The fact that these exponents are
independent of where and at what angle the sequence crosses the
critical curve is a consequence of the scaling laws derived in
Sect.~\ref{sec2}.

Before the predictions (\ref{scaling3}) can be compared with the
numerical data, including all constant factors, we need to determine
the family-dependent parameters $\eta_{*0}$, $C_0$, $C_1$ in
(\ref{reduced}) and $K$ in (\ref{Pdef}).  We first consider the
$\Omega = 0$ sequence, for which (\ref{Mscaling3}) reduces to
\begin{equation}
M \simeq {\bar\eta}^{\gamma_M} = C_0^{\gamma_M} (\eta - \eta_{*0})^{\gamma_M}.
\end{equation}
Fitting this expression to numerical data then yields the parameters
$\eta_{*0} = 1.0183772$ (in agreement with
\cite{EvansColeman}) and $C_0 \simeq 0.28$. Fitting to one of
the rotating families, for example the $\Omega = 0.05$ sequence, then
yields $C_1 \simeq 4.5$ in a similar fashion.  Finally, we insert
(\ref{reduced}) together with the now known coefficients $C_0$ and
$C_1$ into (\ref{Pdef}), set $P = 0$, and fit the resulting relation
between $\eta$ and $\Omega$ to the parabola describing the critical
curve ($\eta_*,\Omega_*)$ to obtain $K \simeq 0.0046$.  We will use
these parameters in all of the following plots, which makes them
heavily overdetermined. 

%%%%%%%%%%%%%%%%%%%%%%%%%%%%%%%%%%%%%%%%%%%%%%%%%%%%%%%%%%%%%%%%%
\begin{figure}
\includegraphics[scale=0.65, angle=0]{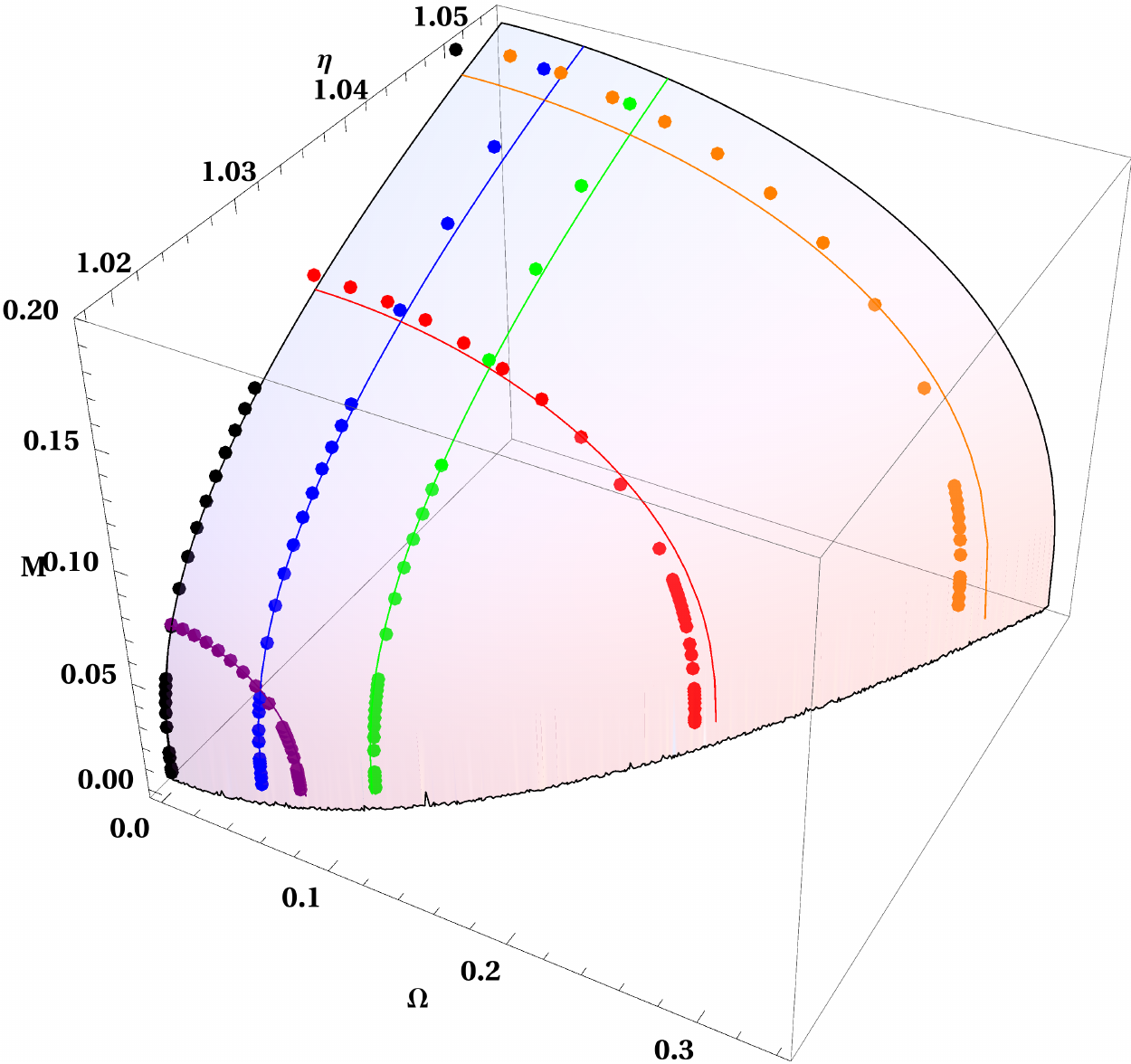} 
\includegraphics[scale=0.65, angle=0]{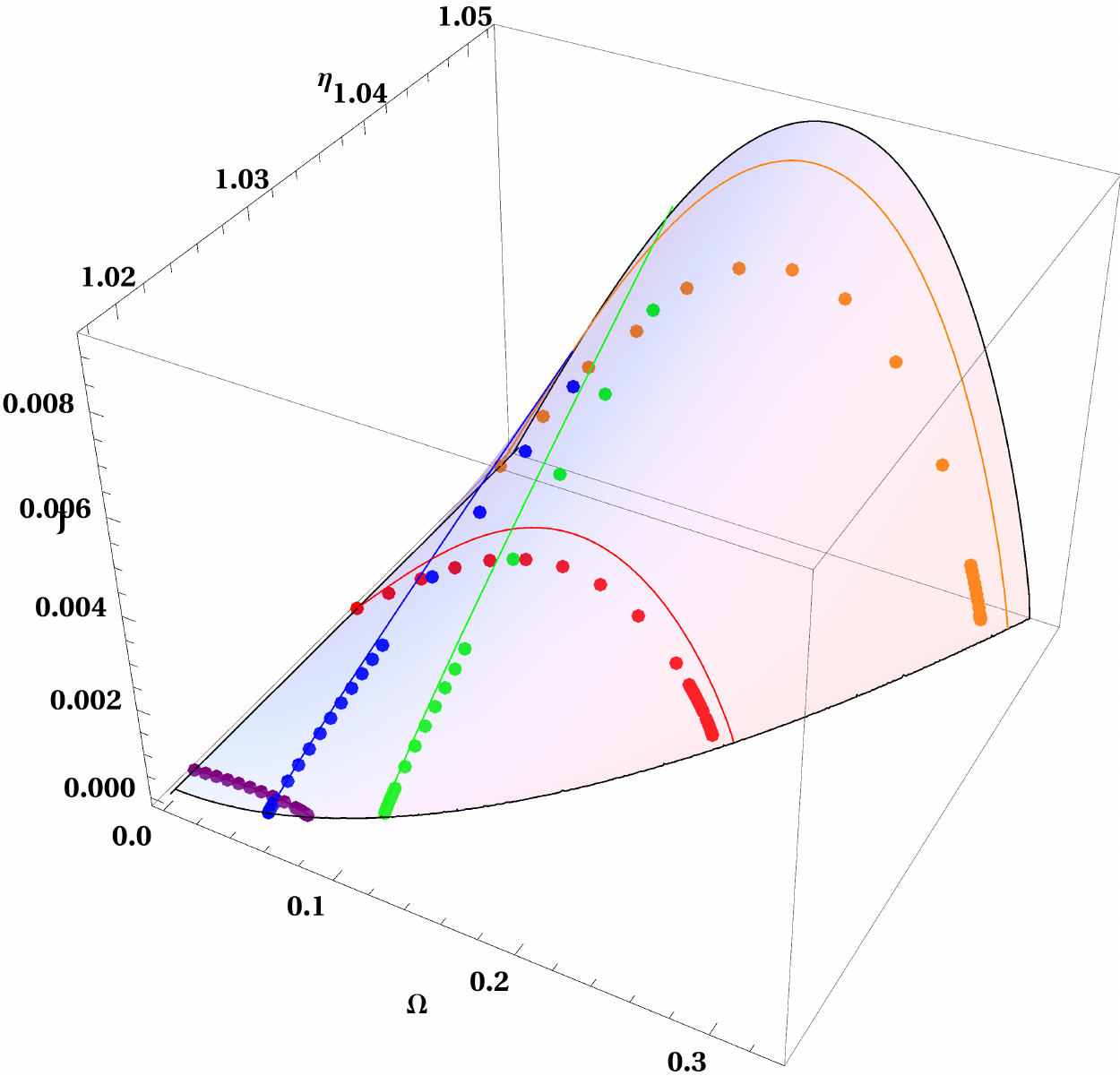} 
\caption{The theoretical predictions (\ref{scaling3}) (continuous
  surface and curves) for black-hole masses $M$ (top) and angular momenta $J$
  (bottom), together with numerical data from six 1-parameter families
  for $\Omega=0,$ 0.05 and 0.1 (black, blue, green) and $\eta=1.02,$
  0.1035 and 1.0505 (purple, red, orange).}
\label{figure:omegaetaMJplot}
\end{figure}
%%%%%%%%%%%%%%%%%%%%%%%%%%%%%%%%%%%%%%%%%%%%%%%%%%%%%%%%%%%%%%%%%

%%%%%%%%%%%%%%%%%%%%%%%%%%%%%%%%%%%%%%%%%%%%%%%%%%%%%%%%%%%%%%%%%
\begin{figure}
\includegraphics[scale=0.65, angle=0]{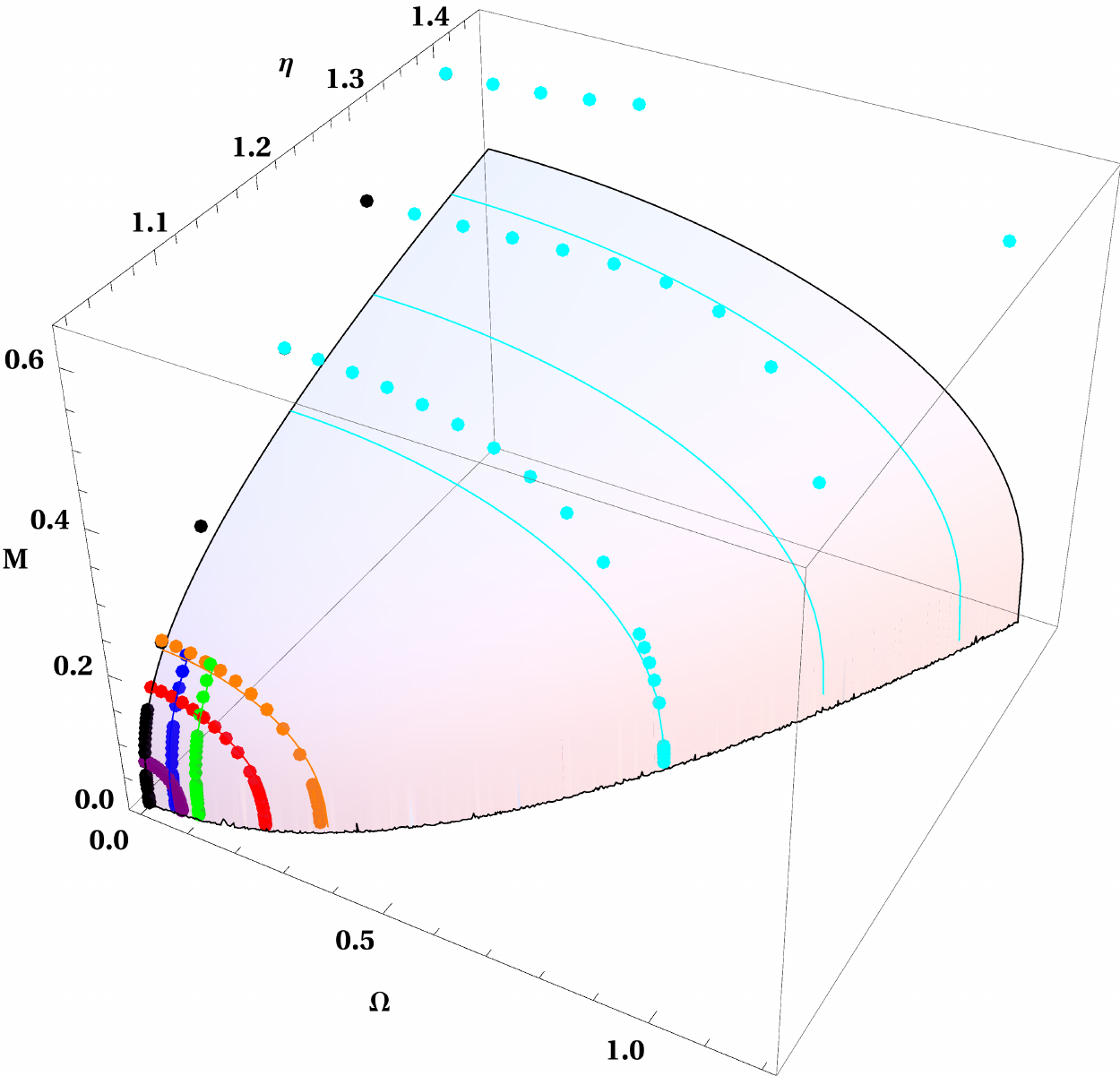} 
\includegraphics[scale=0.65, angle=0]{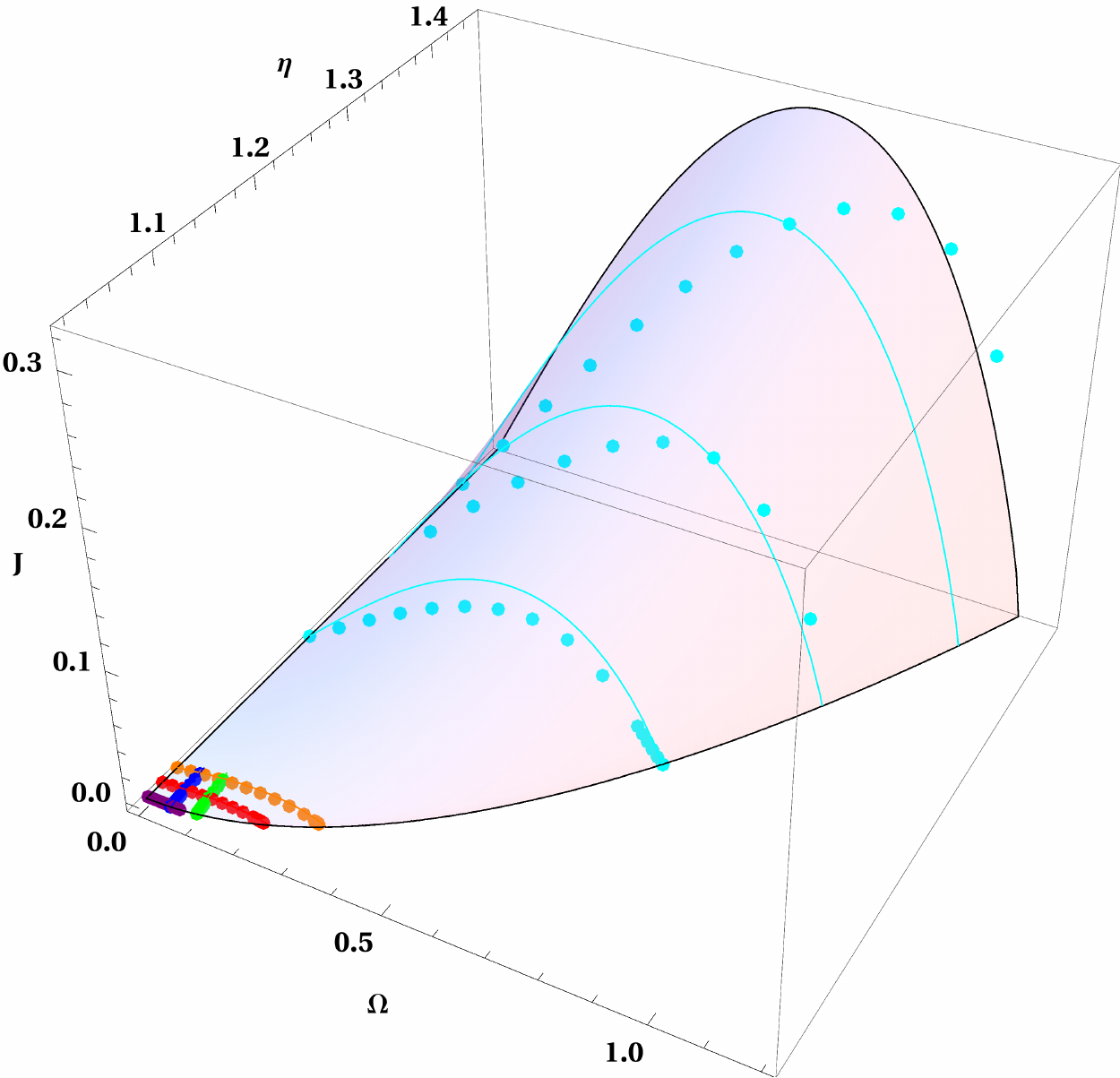} 
\caption{A zoom-out from the previous figure, showing the previous
  families of initial data, and in addition $\eta=1.2$, $1.3$ and
  $1.4$ (all in cyan). These additional data show increasing
  deviations from the theoretical model further away from the critical
  point, as is in fact required to maintain $|J|/M^2<1$.}
\label{figure:omegaetaMJplotlarge}
\end{figure}
%%%%%%%%%%%%%%%%%%%%%%%%%%%%%%%%%%%%%%%%%%%%%%%%%%%%%%%%%%%%%%%%%

In Figs.~\ref{figure:omegaetaMJplot} and
\ref{figure:omegaetaMJplotlarge} we plot the theoretical predictions
for black-hole masses $M$ and angular momenta $J$ in our
two-dimensional parameter space.  Specifically, we plot $M$ and $J$ as
functions of $\eta$ and $\Omega$ as given by (\ref{scaling3}), based
on the parameters determined above. Fig.~\ref{figure:omegaetaMJplot}
contains the data already used in \cite{BaumgarteGundlach}, while
Fig.~\ref{figure:omegaetaMJplotlarge} contains additional data further
away from the critical point.  Fig.~\ref{figure:omegaetaMJplot}
already suggests good agreement between the theoretical predictions
and the numerical data {not too far from the critical point}, but for
a clearer quantitative comparison between model and data we also plot
1-parameter families of either constant $\Omega$ or $\eta$.

For 1-parameter families at constant ${\bar\Omega} = {\bar\Omega}_*$
(which appear as vertical lines in Fig.~1 of \cite{BaumgarteGundlach})
the scaling laws (\ref{scaling3}) can be written as
\begin{subequations} \label{scalingconstq}
\begin{eqnarray}
\label{Mscalingconstq}
M&\simeq&({\bar\eta}-{\bar\eta}_*)^{\gamma_M}, \\
\label{Jscalingconstq}
J&\simeq&({\bar\eta}-{\bar\eta}_*)^{\gamma_J}{\bar\Omega}_*,
\end{eqnarray}
\end{subequations}
where we have used ${\bar\eta}_* = K {\bar\Omega}_*^2$. Inserting the expressions
(\ref{reduced}) for the reduced parameters we may then define the
dimensionless expressions
\begin{subequations} \label{scalingconstqbis}
\begin{eqnarray}
\label{Mscalingconstqbis}
M_\Omega := {M\over
  (C_0\eta_*)^{\gamma_M}}&\simeq&\left({\eta\over\eta_*}-1\right)^{\gamma_M},
\\
\label{Jscalingconstqbis}
J_\Omega := {J\over C_1\Omega_*(C_0\eta_*)^{\gamma_J}}
&\simeq&\left({\eta\over\eta_*}-1\right)^{\gamma_J}
\end{eqnarray}
\end{subequations}
for sequences of constant $\Omega$.

Similarly, for 1-parameter families at constant ${\bar\eta}={\bar\eta}_*$ (which
appear as horizontal lines in Fig.~1 of \cite{BaumgarteGundlach}) the
scaling laws (\ref{scaling3}) can be written as
\begin{subequations} \label{scalingconstp}
\begin{eqnarray}
\label{Mscalingconstp}
M&\simeq&\left[{\bar\eta}_*\left(1-{{\bar\Omega}^2\over{\bar\Omega}_*^2}\right)\right]^{\gamma_M}, \\
\label{Jscalingconstp}
J&\simeq&\left[{\bar\eta}_*\left(1-{{\bar\Omega}^2\over{\bar\Omega}_*^2}\right)\right]^{\gamma_J}{\bar\Omega}.
\end{eqnarray}
\end{subequations}
For sequences of constant $\eta$ we then define
\begin{subequations} \label{scalingconstpbis}
\begin{eqnarray}
\label{Mscalingconstpbis}
M_\eta := {M\over M_{\rm max}}
&\simeq&\left(1-{\Omega^2\over\Omega_*^2}\right)^{\gamma_M},
\\
\label{Jscalingconstpbis}
J_\eta := {J\over J_{\rm max}}
  &\simeq&{1\over
  C_J}\left(1-{\Omega^2\over\Omega_*^2}\right)^{\gamma_J}{\Omega\over\Omega_*}
\end{eqnarray}
\end{subequations}
(these are Eqs.~(9) and (10) of \cite{BaumgarteGundlach}), 
where we have abbreviated
\begin{subequations} \label{MJmax}
\begin{eqnarray}
M_{\rm max} & = & [C_0(\eta_* - \eta_{*0})]^{\gamma_M}, \\
J_{\rm max} & = & C_J C_1 \Omega_* [C_0 (\eta_* - \eta_{*0})]^{\gamma_J}.
\end{eqnarray}
\end{subequations}
To leading order, these quantities are the maximum mass and angular
momentum along the sequence of constant $\eta$, and $C_J\simeq 0.4025$
is the maximum of $x(1-x^2)^{\gamma_J}$ on the interval $[0,1]$.

%%%%%%%%%%%%%%%%%%%%%%%%%%%%%%%%%%%%%%%%%%%%%%%%%%%%%%%%%%%%%%%%%
\begin{figure}
\includegraphics[scale=0.6, angle=0]{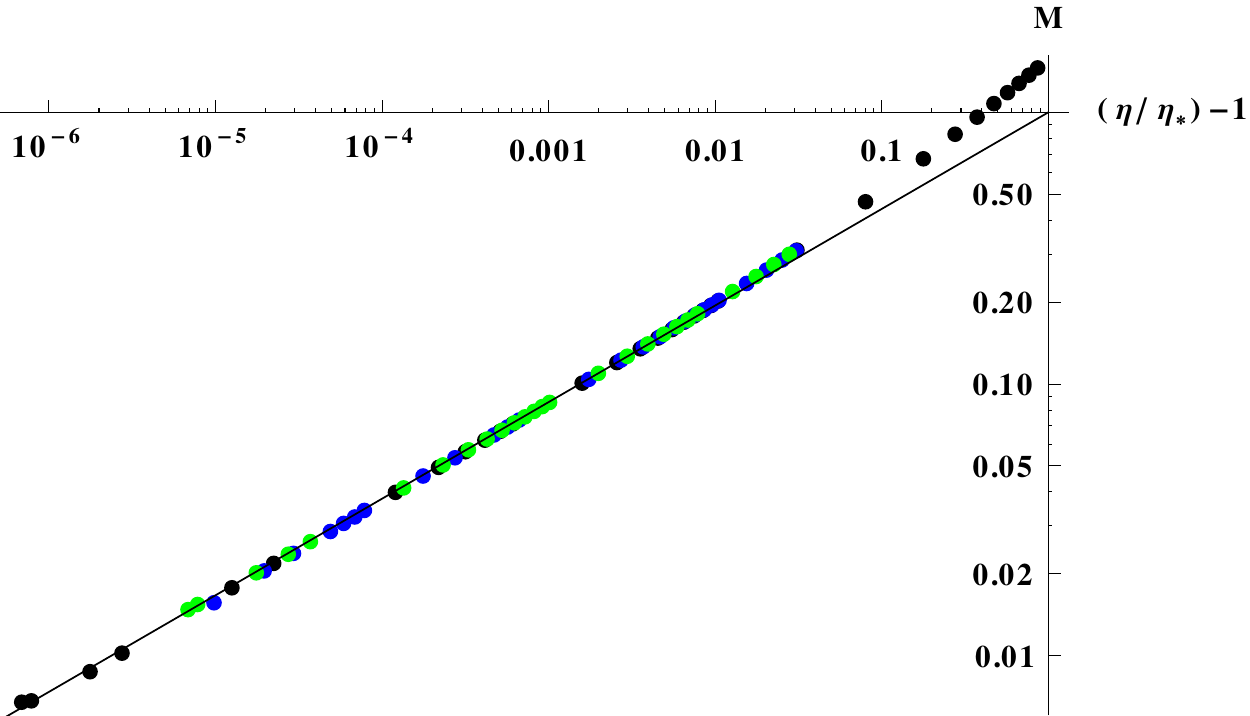} 
\includegraphics[scale=0.6, angle=0]{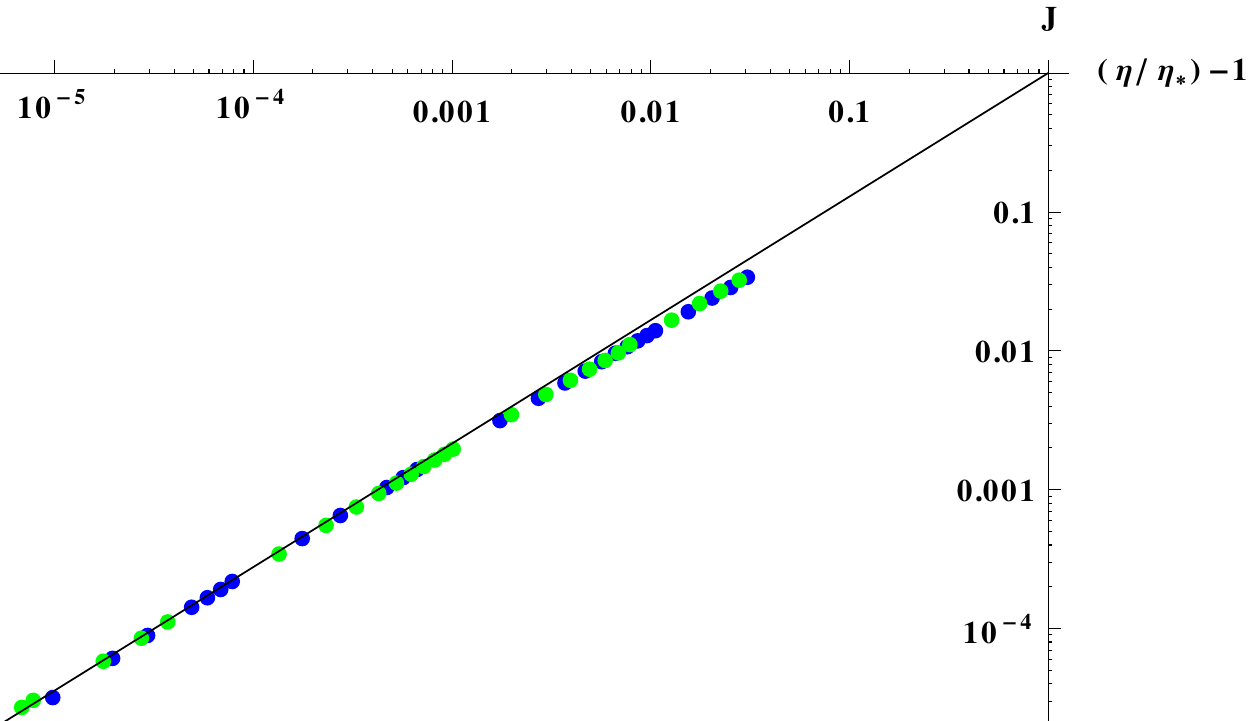} 
\caption{Verification of (\ref{Mscalingconstqbis}) for $M$ (top) and
  (\ref{Jscalingconstqbis}) for $J$ (bottom). The continuous curves are
  the analytic expressions on the right-hand sides of these equations,
  plotted against $(\eta/\eta_*)-1$. The dots are numerical values for
  the left-hand sides, for the three 1-parameter families of initial
  data at constant $\Omega=0,0.05,0.1$ (black, blue, green, as for the
  same data in Fig.~\ref{figure:omegaetaMJplot}). The critical surface
  is at an infinite distance to the bottom left in this log-log plot. There is
  obviously no $J$ plot for the $\Omega=0$ family.}
\label{figure:MJvsetaplots}
\end{figure}
%%%%%%%%%%%%%%%%%%%%%%%%%%%%%%%%%%%%%%%%%%%%%%%%%%%%%%%%%%%%%%%%%

%%%%%%%%%%%%%%%%%%%%%%%%%%%%%%%%%%%%%%%%%%%%%%%%%%%%%%%%%%%%%%%%%
\begin{figure}
\includegraphics[scale=0.6, angle=0]{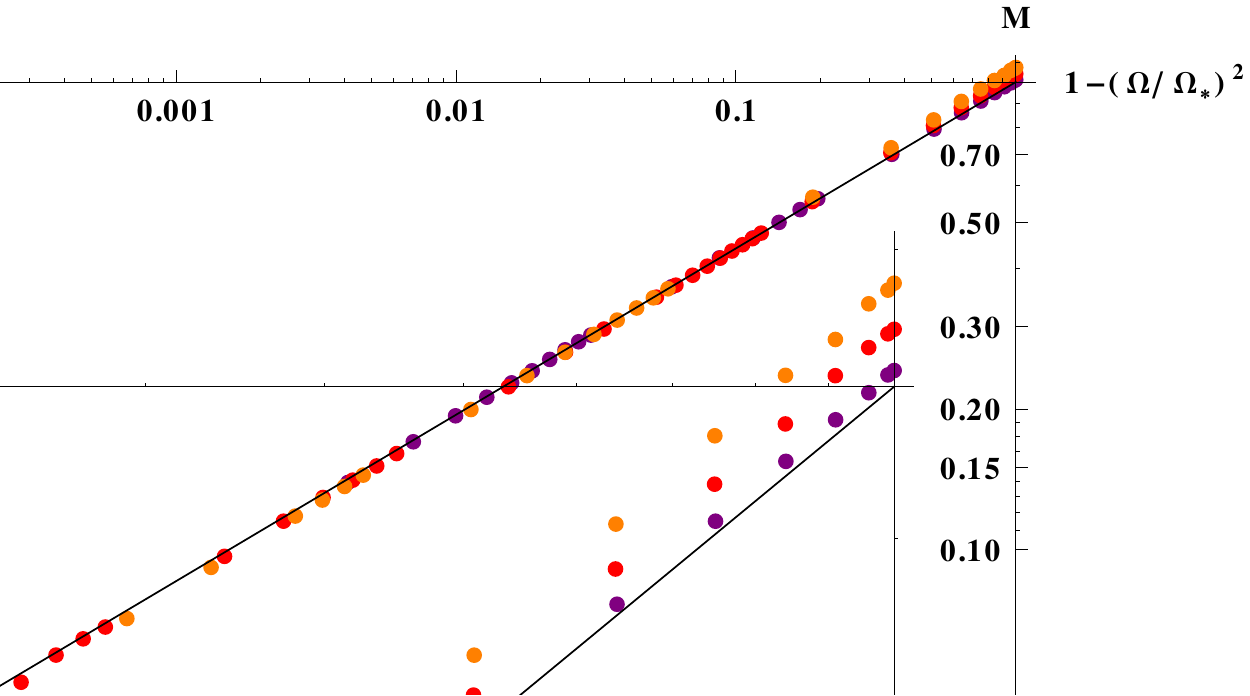} 
\includegraphics[scale=0.6, angle=0]{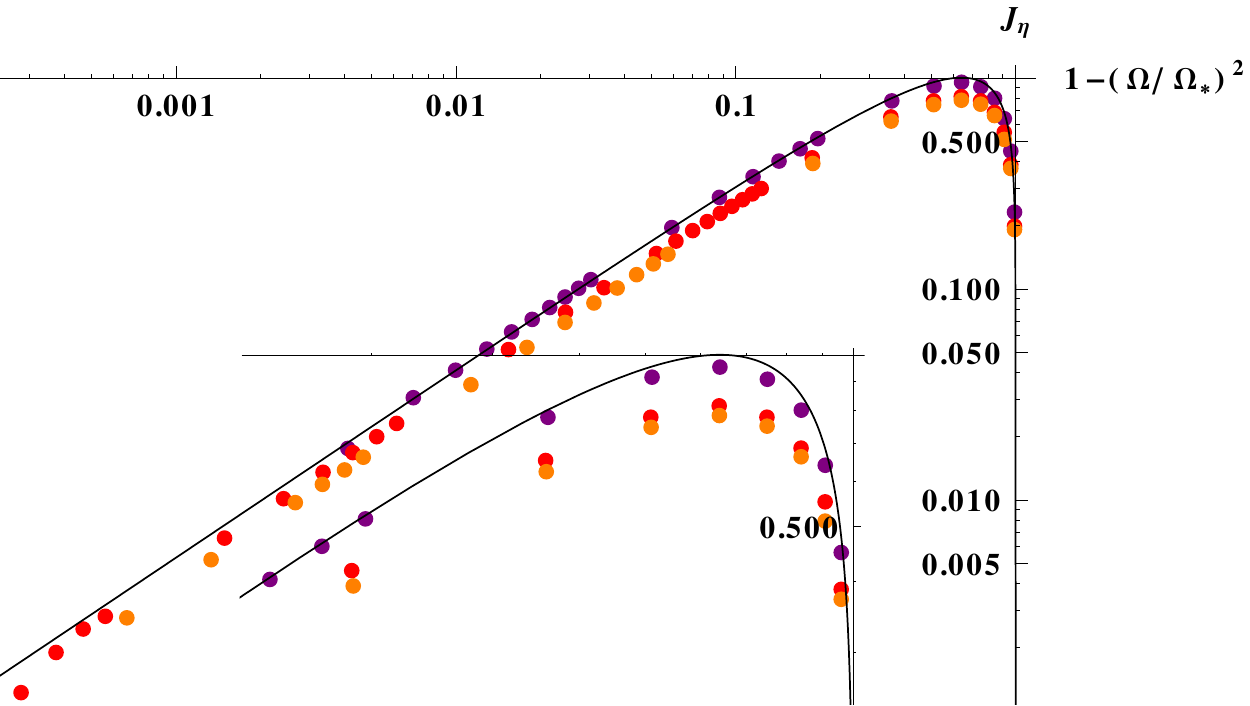} 
\caption{Verification of (\ref{Mscalingconstpbis}) for $M$ (top) and
  (\ref{Jscalingconstpbis}) for $J$ (bottom). The continuous curves are
  the analytic expressions on the right-hand sides of these equations,
  plotted against $1-(\Omega/\Omega_*)^2$. The dots are numerical
  values for the left-hand sides, for the three 1-parameter families
  of initial data at constant $\eta=1.02,0.1035,1.0505$ (purple, red,
  orange, as for the same data in
  Fig.~\ref{figure:omegaetaMJplot}). The critical surface is at
  at an infinite distance to the bottom left in this log-log plot.
  $\Omega=0$ is at the right edge of the plot, and the fall-off there
  represents $J\sim\Omega$ for small $\Omega$ in this log-log plot.}
\label{figure:MJvomegaplots}
\end{figure}
%%%%%%%%%%%%%%%%%%%%%%%%%%%%%%%%%%%%%%%%%%%%%%%%%%%%%%%%%%%%%%%%%

In Fig.~\ref{figure:MJvsetaplots} we plot the masses and angular
momenta of black holes formed from supercritical data in families of
constant $\Omega$, and in Fig.~\ref{figure:MJvomegaplots} for families
of constant $\eta$.  In these graphs, the solid lines represent the
right-hand sides of Eqs.~(\ref{scalingconstqbis}) and
(\ref{scalingconstpbis}), while the points mark the numerical data,
rescaled as on the left-hand sides of (\ref{scalingconstqbis}) and
(\ref{scalingconstpbis}).  These rescaling factors include the
parameters $C_0$, $C_1$ and $\eta_{*0}$ as determined above for the
entire two-parameter family of initial data; i.e.~these parameters are
not fitted separately to the individual 1-parameter families shown in
the figures.  In addition, the rescaling factors include the critical
values $\eta_*$ or $\Omega_*$ for each 1-parameter family.  In
principle, these could be computed by setting
$P=0$ in (\ref{Pdef}), which yields
\begin{equation} \label{crit_curve}
C_0 ( \eta_* - \eta_{*0}) \simeq K (C_1 \Omega_*)^2.
\end{equation}
Using this expression would result in a small error in the critical
parameters but a diverging {\em relative} error for $M$ and $J$ at the
collapse threshold, thus hiding the physically correct scaling law.
This is already familiar elsewhere in the numerical study of critical phenomena:
numerical values of, say, a critical exponent or the critical
parameter of a 1-parameter family converge to a continuum limit with
increasing numerical resolution, but data at a given resolution will
show critical scaling only for the critical parameters obtained for
that resolution, rather than higher-resolution values.  For a similar
reason, we fit $\eta_*$ or $\Omega_*$ for the individual 1-parameter
families, rather than using (\ref{crit_curve}), so that the data
reveal critical scaling even close to the black-hole threshold.

For data sufficiently close to the critical point, there is good
agreement between the prediction and numerical values. This is shown
in Fig.~\ref{figure:omegaetaMJplot}.  The differences increase further
away from the {critical point}, where we expect increasing deviations
between the leading-order scaling laws and non-linear numerical
evolutions. In particular, the maxima of $|\delta|$ with respect to
${\bar\Omega}$ at constant ${\bar\eta}$ lie on the parabola
\begin{equation}
\label{maxJoM2curve}
{\partial\over\partial{\bar\Omega}}\delta({\bar\eta},{\bar\Omega})=0
 \quad\Rightarrow\quad
{\bar\eta}\simeq (1-2\epsilon)K{\bar\Omega}^2.
\end{equation}
Along this curve, parameterised by ${\bar\eta}$, our model predicts
\begin{equation}
\max_{\bar\Omega} {J\over M^2}({\bar\eta},{\bar\Omega})\simeq K^{-{1\over
    2}}(-2\epsilon)^{-\epsilon}(1-2\epsilon)^{\epsilon-{1\over2}}{\bar\eta}^{{1\over2}-\epsilon},
\end{equation}
which increases monotonically with ${\bar\eta}$. Hence we expect our
model to break down well before it predicts {$|J|/M^2>1$}. Indeed, we
find that further away from the critical point the numerically found
masses are larger than predicted, and the angular momenta smaller than
predicted, such that $J/M^2$ increases but never goes beyond
unity. These observations are demonstrated in
Fig.~\ref{figure:omegaetaMJplotlarge}.

We would like to emphasize a difference between
Figs.~\ref{figure:MJvsetaplots} and \ref{figure:MJvomegaplots} and and
the scaling plots of \cite{BaumgarteGundlach}. In Fig.~2 of
\cite{BaumgarteGundlach} we fitted power-laws to individual
1-parameter families, whereas here we compare the numerical data with
predictions for the entire two-parameter family, given the constants
$C_0$, $C_1$ and $\eta_{*0}$. Similarly, for Fig.~4 of
\cite{BaumgarteGundlach}, which shows the same data as
Fig.~\ref{figure:MJvomegaplots} here, we computed $M_{\rm max}$ and
$J_{\rm max}$ in (\ref{scalingconstpbis}) from the numerical data
(which did not require knowledge of the constants $C_0$, $C_1$ and
$\eta_{*0}$), whereas here we compute $M_{\rm max}$ and $J_{\rm max}$
from (\ref{MJmax}), using the constants $C_0$, $C_1$ and $\eta_{*0}$.
The agreement found here is therefore a more stringent comparison
between theoretical predictions and numerical data than that presented
in \cite{BaumgarteGundlach}.

%%%%%%%%%%%%%%%%%%%%%%%%%%%%%%%%%%%%%%%%%%%%%%%%%%%%%%%%%%%%%%%%%%%%%%%%

\section{Summary and Discussion}
\label{sec4}

%%%%%%%%%%%%%%%%%%%%%%%%%%%%%%%%%%%%%%%%%%%%%%%%%%%%%%%%%%%%%%%%%%%%%%%%

We have extended arguments previously presented in
\cite{angmom,critfluidpert,scalingfunctions} to derive new closed-form
scaling laws for critical collapse with angular momentum.  The main
result, Eq.~(\ref{scaling1}), in principle involves scaling functions
$F_{M,J}$ that need to be determined numerically, but adopting the
lowest-order expansions (\ref{trivialscalingfunctions}) of these
scaling functions yields the predictions (\ref{scaling3}).  We have
compared these predictions with numerical simulations for rotating
radiation fluids and find excellent agreement.  We believe that the
agreement between theory and numerics for the mass and angular
momentum scaling, even so far from critical collapse that there is no
clear sign of an intermediate self-similar regime, is another example
of the ``unreasonable effectiveness of perturbation theory'', similar
to post-Newtonian approximations \cite{Will} and the close-limit
approximation \cite{PricePullin} for binary black hole mergers.

This paper also demonstrates the necessary interplay between theory
and numerical experiment in critical collapse.  In the previous
treatment \cite{angmom,critfluidpert,scalingfunctions} CG implicitly
truncated the expansion (\ref{Pdef}) as $P=\bp$, which
renders the black-hole threshold as a straight line. The numerical
results of \cite{BaumgarteGundlach} showed that the resulting picture
is qualitatively incorrect. In hindsight,
$\order(\bp)=\order(\bq^2)$, so that the consistent
lowest-order truncation of the coefficients $P$ and $Q$ is
(\ref{PQdef}), and the black-hole threshold is a parabola.  For the
case of one unstable mode, CG correctly predicted the scaling
behaviour (\ref{scaling2}) for small $\bq$, but by implicitly setting
$K=0$ (in the notation of this paper) he missed the equally
interesting behavior of (\ref{scaling3}) for finite $\bq$, in
particular (\ref{scalingconstp}). The formal prediction of nontrivial
universal scaling functions (\ref{scaling1}) for the mass and angular
momentum has not changed, but $\delta$ is now given in terms of $P$
and $Q$, not $\bp$ and $\bq$, with the lowest-order consistent
truncation given by Eq.~(\ref{PQdef}).\footnote{The similarity between
  our Fig.~\ref{figure:omegaetaMJplot} and Fig.~2 of
  \cite{scalingfunctions} is, unfortunately, accidental.}

In Sect.~\ref{sec3} we compared the theory with numerical results the
perfect fluids with the equation of state $P = \kappa \rho$ where
$\kappa = 1/3$ (the radiation fluid).  As discussed in the
Introduction, for $1/9 < \kappa \lesssim 0.49$ the critical solution
has only one unstable mode, so that $\epsilon < 0$ in particular for
$\kappa=1/3$.  Eq.~(\ref{definedelta}) then implies $\delta
\rightarrow 0$ as the black-hole threshold $P = 0$ is approached,
which, by (\ref{JM2}), implies $J/M^2 \rightarrow 0$, in agreement
with our numerical findings.  

It is precisely the fact that $\delta \rightarrow 0$ at the black-hole
threshold that allows us to use the simple expansions
(\ref{trivialscalingfunctions}) for the scaling functions in
(\ref{scaling2}).  For $\kappa < 1/9$, on the other hand, we expect
two unstable modes, so that $\epsilon > 0$, in which case
(\ref{definedelta}) suggesst that $\delta$ grows without bound as $P
\rightarrow 0$.  In this case, therefore, use of the expansions
(\ref{trivialscalingfunctions}) can no longer be justified. We plan to
explore this regime in the future.

We would like to stress that the derivation of (\ref{scaling3}) relies
on at least four logically independent perturbation approximations:
(i) Sufficiently near the collapse threshold the time evolution
actually goes through a phase where it can be described by the
critical solution plus linear perturbations. (ii) In at least part of
this phase, all decaying linear perturbations, except possibly for
$Z_1$, can be neglected. (iii) The scaling functions can be
approximated by their leading orders
(\ref{trivialscalingfunctions}). (iv) The amplitudes $P$ and $Q$ can
be approximated by their leading orders (\ref{PQdef}).

It is hard to know which of these approximations causes the most
significant deviations from (\ref{scaling3}) as we move away from the
critical point. One might consider adding higher powers of $\bp$ and
$\bq$ to the expressions for $P$ and $Q$ as indicated in (\ref{PQdef})
to control (iv). However, as $M(p,q)$ and $J(p,q)$ are smooth
functions away from the critical surface, we can always make our model
fit the data precisely by fitting $P(p,q)$ and $Q(p,q)$, and so this
fit has no additional predictive power.

One might also consider extending (\ref{trivialscalingfunctions}) to a
power series in $\delta$ to control (iii). However, we can already fit
the model to the numerics perfectly by fitting $P$ and $Q$, even if we
set $F_M=1$ and $F_J=\delta$, and so we cannot determine these
universal functions by such a fit. They could, however, be defined
directly by the time evolution of the two $1$-parameter family of
universal intermediate initial data (\ref{data}). The scaling
functions will play a non-trivial role when there are two unstable
modes, as $\delta\to 0$ is then still realised as $\bq\to 0$ but
$\delta\to\infty$ is realised as $\bq\to \bq_*(\bp)$.

We note in closing that Aguilar-Martinez \cite{Aguilar} has presented
preliminary numerical results for the critical collapse, in {\em
  Newtonian} gravity, of an axisymmetric rotating fluid with equation
of state $P\propto \rho_0^\Gamma$ (where $\rho_0$ is the rest mass
density) with $\Gamma=10^{-5}$. The critical solution appears to have
two unstable modes in this case.  Aguilar-Martinez gives theoretical
values of $\lambda_0\simeq 9.4643$ and $\lambda_1=1/3$, implying
$\epsilon\simeq 0.3522$.  The black-hole threshold found in his
simulations appears to be well approximated by a
parabola. Interestingly, $M(p,q)$ and $J(p,q)$ also approach zero at
the black-hole threshold even though there are two unstable modes. It
will be interesting to compare the case of two unstable modes in
general relativity.

%%%%%%%%%%%%%%%%%%%%%%%%%%%%%%%%%%%%%%%%%%%%%%%%%%%%%%%%%%%%%%%%%%%%%%%%

\acknowledgements

This work was supported in part by NSF grant PHY-1402780 to Bowdoin College.

%%%%%%%%%%%%%%%%%%%%%%%%%%%%%%%%%%%%%%%%%%%%%%%%%%%%%%%%%%%%%%%%%%%%%%%%

%%%%%%%%%%%%%%%%%%%%%%%%%%%%%%%%%%%%%%%%%%%%%%%%%%%%%%%%%%%%%%%%%%%%%%%%

%%%%%%%%%%%%%%%%%%%%%%%%%%%%%%%%%%%%%%%%%%%%%%%%%%%%%%%%%%%%%%%%%%%%%%%%

\end{document}